# Nuclear Tracks from Cold Dark Matter Interactions in Mineral Crystals: A Computational Study.

J. I. Collar [§] and F. T. Avignone III

Department of Physics and Astronomy
University of South Carolina
Columbia, SC 29208

Recoiling nuclei from Cold Dark Matter (CDM) elastic scattering interactions with the constituent elements of some minerals may produce etchable damage tracks in the crystal structure. Present calculations show that in muscovite mica, CDM tracks from recoiling potassium atoms could be readily distinguished from others due to alpha-decays in the uranium and thorium chains. Under favorable conditions, this technique could greatly improve the existing limits on Weakly Interacting Massive Particles as the constituents of the galactic halo.

[§]Corresponding author. e-mail: ji.collar@scarolina.edu

# 1. Introduction.

Numerous Dark Matter candidates have been proposed, that could constitute the approximately 90% of the universe unobservable by conventional astronomy. Ample literature exists, covering various members of the candidate arena [1,2]. Experimental searches are currently operational or under development [1]; Baryonic Dark Matter (BDM), in the form of Massive Astrophysical Compact Halo Objects (MACHOs) can account for part of the galactic dark halo. The detection of MACHOs (in the range $10^{-7}$ to $10^{-1}$ solar masses) may be possible by means of gravitational micro-lensing. Preliminary reports [3] indicate that such events may have been observed. While the data at hand are still inconclusive, there seems to be an initial agreement between the frequency of these observations and what is expected from an important contribution of BDM to the halo mass. Nevertheless, even if BDM in this form has been observed, it cannot be the single source of dark matter in the Universe without contradicting the predictions of Primordial Nucleosynthesis.

Hot Dark Matter, light relativistic neutrinos, is another possibility probed by experiments designed to search for a finite neutrino mass ($m_\nu \approx$ few eV's). Probably the most populated candidate group is that of Cold Dark Matter (CDM), consisting of heavy, non-relativistic particles, extending over a large range of possible masses and coupling constants. As a subdivision of the CDM category, Weakly Interacting Massive Particles (WIMPs) comprise some of the most interesting candidates. The theoretical justification for the existence of WIMPs stems from extensions of the Standard Model of the electro-weak interaction, and in some cases, have the added attraction of having been initially postulated to solve theoretical problems unrelated to that of Dark Matter, as it is the case for the Lightest Supersymmetric Partner (LSP, generally considered to be a neutralino, product of the predictions of the Minimal Supersymmetric Standard Model). An important reason why WIMPs have become so prevalent in the dark matter theater is their natural ability to account for the dark galactic halo; particles with masses in the range few GeV - few TeV and interactions of typical weak strength can explain the formation of structures at the galactic scale and have a large enough relic density to account for the halo mass necessary to explain the observed galaxy dynamics.

The WIMP hypothesis has prompted interest in the development of new detection techniques and the improvement of existing ones; their aim, the direct detection of CDM signatures, specifically the recoil energy from coherent elastic scattering off nuclei [4]. For this recoil energy to be measurable, and because of the small expected interaction rates, dramatic improvements in the radiopurity of the materials surrounding the detectors have been achieved; clever techniques have been devised to separate the expected WIMP signal from some of the remaining backgrounds, and new technologies have been developed. A complete review of these experimental efforts can be found in Ref. 1.

Solid State Nuclear Track Detectors (SSNTDs) [5,6] have been recently proposed as an alternate method of detection [7]. These electrical insulators (or poor semiconductors) have found many interdisciplinary applications, including searches for magnetic monopoles [8,9]. Heavily ionizing particles ($\alpha$-particles, fission fragments, recoiling nuclei, etc.) undergo elastic nuclear collisions and electronic energy losses. These result in tracks of latent solid-state damage (displaced lattice atoms, broken molecular bonds, etc.) that can later be revealed by etching with a chemical reagent. Essentially, etching occurs along a particle track faster than along the undamaged surface of the bulk material, due to the rapid dissolution of the damaged region. A surface etch-pit is thus formed and can be inspected by optical or electron microscopy, among other techniques. The main characteristic of interest in SSNTD materials is their insensitivity to minimum ionizing radiation (i.e., gamma and x-rays, ß-particles, etc.), which constitute the largest source of background competing with the WIMP signal in other searches.

## 2. Muscovite Mica: a Natural SSNTD.

SSNTDs can be divided into two groups: a) man-made polymers and plastics (CR-39, Lexan, etc.) with high sensitivities and low densities of latent tracks from the uranium and thorium chains and, b) naturally occurring materials (minerals and glasses), of which muscovite mica seems to be the most sensitive [6]. The sensitivity is related to the threshold in stopping power of the incident particle, below which no etchable tracks can be formed. Several models explain the formation of tracks, but none defines this registration threshold in all cases. For each material and irradiation, precise calibrations must be made to relate the appearance of etchable tracks to the velocity of the projectile, its number of primary ionizations close to the particle path, restricted energy loss, nuclear and electronic stopping powers, etc. There are excellent books on the subject by Fleischer, Price and Walker [6] and more recently by Durrani and Bull [5] .

In the absence of thermal annealing, damaged regions can remain unaltered almost indefinitely, thus cumulatively storing the signature of many particle interactions for periods of up to billions of years. This feature has made possible the dating of archeological artifacts, rocks and meteorites ("fission-track dating") [6]. This property can be exploited to show that recoiling nuclei from elastic scattering of WIMPs in some SSNTD ancient minerals may have left enough cumulative damage for a characteristic population of etch-pits to be observable and distinguishable from those arising from other known processes. Even though the composition of these minerals would not make them especially promising as CDM detectors because of their low expected rate of interaction, the long exposure to the CDM flux over the lifetime of the material amply compensates for it.

Muscovite mica ( $K_2Al_4[Si_6Al_2O_{20}]-(OH,F)_4$; $\rho_{Mica} \sim 2.8 \text{ g}/\text{cm}^3$) is the most extensively studied of natural SSNTDs. Irradiations of mica sheets with low-velocity heavy ions ($8 \leq Z \leq 90$) [10] have produced etchable tracks for $0.0005\,c < v < 0.0025\,c$, a regime similar to the expected nuclear recoil energies from WIMP scattering (T ~ tens of keV). More precise measurements are under way [11]. From these results, Price and Salamon [8] extracted an expression for the etching velocity along the damage-track when the reagent is 40% Hydrofluoric acid at 25° C :

$$V_t \cong 0.012 \cdot S_n \;(\mu m/hour), \qquad (1)$$

where $S_n$ is the nuclear stopping power of an ion of given mass and velocity, expressed in units of GeV cm$^2$ /g . For these low velocities, the electronic stopping power $S_e$ is a small fraction of the total stopping power ( $S_t = S_e + S_n$ ) and its contribution was neglected in Eq. (1). Under the same etching conditions (which apply in the rest of this paper), the rate of chemical attack perpendicular to the cleavage surface in the absence of a track is measured to be $V_\perp = 0.027$ (µm / hour). Geometric arguments imply that a particle penetrating a surface at a dip angle θ, forms an etch-pit only when $\theta > \theta_c$, where

$$\theta_c = \arcsin(V_\perp / V_t) \qquad (2)$$

(otherwise the surface recedes too fast for the etch-pit to materialize) [5,6]. The depth of the etch-pit, $\zeta$, is given by $\zeta = (V_t \sin\theta - V_\perp) \cdot t$, where t is the etching time. When the product $V_t t$ exceeds the range of the particle, the pit can remain visible and $\zeta$ constant for much longer etching times, whereas the diameter of the pit continues to increase [5,12]. This implies a monotonic increase in the number of observed etched tracks with t. This linearity, however, has been observed to break down at very large values of t, as many of the early pits, especially shallow ones, start to lose their sharpness.

Under these etching conditions, three distinct groups of pits are revealed in unirradiated muscovite mica [13], differing greatly in their values of $\zeta$ but all of them originating from the U and Th chains. The first group, with $<\zeta> \sim 20$ μm, is due to nuclear fragments originating in the spontaneous fission of $^{238}$U, where $S_t \sim 25$ GeV cm$^2$/g. The second is a dense collection of very shallow pits ($\zeta < 0.02$ μm) that has its origin in the recoiling nuclei from α-emission in $^{238}$U and $^{232}$Th [14]. The surface density ratio between these small pits ($S_t \sim 15$ GeV cm$^2$/g) and the first group is roughly constant for non-annealed samples, with an average value of $\rho_{rec}/\rho_{SF} = 3.5 \times 10^3$. This is due to the approximately constant U / Th ratio observed in nature (~0.25). Only recently, a third group of intermediate length tracks was found (0.25 μm $< \zeta < 1.5$ μm), visible under an optical microscope with the aid of an interferometric attachment. Price and Salamon [13] identified the origin of such tracks as reactions of 8.8 MeV α-particles from $^{212}$Po in the $^{232}$Th chain, with Al and Si nuclei, leading to Si and P recoil nuclei with kinetic energies ~500 keV and $S_t \sim 3$ GeV cm$^2$/g. Since the stability of latent tracks against thermal annealing is proportional to the degree of damage induced [5] and therefore to $S_t$, the ratio of these faint intermediate length tracks to those from spontaneous fission ($0.03 < \rho_{int}/\rho_{SF} < 0.3$) varies with the thermal history of the sample. It was also shown in Ref. 13 that for micas maintained at normal ambient temperature, this ratio should be at least ~0.15, so that the preservation of faint tracks of $S_t \sim 3$ GeV cm$^2$/g is guaranteed over geological time-scales (~$10^9$ y).

## 3. Ancient Mica as a WIMP Detector.

Figure 1 displays the calculated values of $S_n$ and $S_t$ in muscovite mica for its constituent atoms, with the addition of Phosphorus. Hydrogen is excluded; it has a negligible WIMP cross section. All stopping powers, particle ranges and lateral and longitudinal straggling in this work are derived from the 1992 version of TRIM (Transport of Ions in Matter), a widely used computer code with an accuracy of a few percent in the low-velocity regime [15]. The horizontal line in the figure corresponds to the minimum stopping power for a low velocity ion to leave an etchable track, i.e., in the case when $V_t = V_\perp$. Applying Eq. (1) to potassium as an example, recoils with energies 3 keV < T < 120 keV would produce etchable signatures.

Equation (1) fails to explain the formation of the intermediate length etch-pit tracks; it is evident in Fig. 1 that if their origin is in recoiling Si and P atoms with T ~ 500 keV (which was demonstrated by He-ion calibrations [13]), the single contribution of nuclear recoil energy losses, $S_n$, can not explain the appearance of pits in the $S_t$~3 GeV cm$^2$ g$^{-1}$ regime. Equation (1) is then overly conservative in neglecting the contribution of electronic energy losses to the damage along the ion's trajectory. This observation is supported by recent measurements of the response of CR-39 to low-velocity ions [16], where the relative contribution of $S_e$ to $V_t$ was found to be much larger than expected. The suggested explanation is that while the secondary ions produced in the nuclear collision cascades typical of $S_n$ spread their energy over a range of thousands of angstroms, the secondary (delta) electrons arising from $S_e$ concentrate their energy loss within tens of angstroms of the original ion's path. The etching rate is presumably proportional to the density of damage, and hence the relative importance of $S_e$.

In any case, we observe that for WIMP-nucleus elastic scattering, of the natural elements in mica, only K and possibly Si have stopping powers large enough to form etchable tracks. This is due to the kinematic upper limit on the maximum recoil energy from the collision of a WIMP of mass $m_\delta$ and a nucleus of mass M, which is given by $M v_{max}^2$ (in the limit $m_\delta$>>M) . The speed $v_{max}$ is given by the sum of the speed of the Earth through the galactic halo ($v_{Earth}$ ~ 260 km / s) and the galactic escape velocity ($v_{esc}$ ~ 500 km / s) [2]. As a result, CDM induced recoils of potassium cannot exceed ~230 keV and ~170 keV for silicon; the values are smaller for the rest of mica's constituents.

In light of the previous argument, and in the current absence of specific calibrations for low-velocity K and Si ions in mica, in our analysis we shall examine the two extreme possibilities: a) Eq. (1) being correct, and b) its substitution by the ansatz

$$V_t = 0.012 \cdot (S_n + S_e) \; (\mu m / h). \tag{3}$$

This expression exceeds Eq. (1) by only a few percent for low T, where most of the CDM signal is expected, due to an exponentially decreasing WIMP-nucleus differential rate of scattering, given by [17]:

$$\frac{dR}{dT} = J \, n \int_0^{V\max} f(v) \, v \, \frac{d\sigma}{dT} \, dv. \tag{4}$$

In Eq. (4), J is the number of target nuclei in mica, n is the number density of CDM particles in the galactic halo ($n = \rho_{halo} / m_\delta$), and f(v) is the CDM speed distribution in the Earth's reference frame, generally taken to be a Maxwellian with a dispersion velocity ~ 300 km /s. We have chosen heavy neutrinos as the WIMP candidate for our calculations. Their elastic scattering differential cross section is given by [4]

$$\frac{d\sigma}{dT} = \frac{G_F^2}{8\pi} \cdot \left(\frac{G_f}{G_w}\right)^2 \cdot \left(N - \left(1 - 4 \cdot \sin^2 \theta_w\right) \cdot Z\right)^2 \cdot \frac{m_R^2}{T_{max}} \cdot F(q^2), \tag{5}$$

where $m_R$ is the reduced mass, $T_{max}$ is the maximum recoil energy, N and Z are the number of neutrons and protons in the target nucleus, and $\theta_w$ is the weak mixing angle. $G_F^2$ is the Fermi weak coupling constant ($G_F^2 \cong [290 \text{ GeV}]^{-4} = 5.24 \cdot 10^{-38} \text{ cm}^2$) and the parameter $(G_f / G_w)^2$ allows for coupling constants different from $G_F^2$; $G_f < G_w$ for sub-$Z°$ couplings and $G_f = G_w$ for heavy Dirac neutrinos. The term $F(q^2)$ is a form factor to account for the loss of nuclear coherence for very massive projectiles [17], favoring forward scattering and lower values of T.

The observable quantity in a SSNTD search for CDM is the surface density of WIMP-induced etch-pits revealed after an etching time t, as a function of their depth or diameter. For a random distribution of interaction sites in the sample, and for an initial surface cleaved from well within the body of the material (i.e., a 4π-geometry), this surface density is given by [5]

$$\rho_i^{4\pi} = \frac{1}{2} n_i \left\{ R_a \cos^2 \theta_c + V_\perp t (1 - \sin \theta_c) \right\}, \qquad (6)$$

where $n_i$ is the number of WIMP-nucleus recoils "stored" per unit volume of mica (coming from the ith target element) and $R_a$ is the range of the recoiling nucleus in the mineral. The preservation of pits after their particle tracks have been fully etched is contained in the second term of Eq. (6) ("prolonged-etching factor"). Assuming thermal stability ($\rho_{int} / \rho_{SF} > 0.15$) over the fission-track age of the crystal, A, we have

$$n_i \left[ \frac{\text{tracks}}{\text{cm}^3 \text{ keV}} \right] = A[\text{yr}] \cdot \left. \frac{dR}{dT} \right|_i \left[ \frac{\text{recoils}}{\text{keV kg yr}} \right] \cdot f_i \cdot \rho_{\text{Mica}} \left[ \frac{\text{kg}}{\text{cm}^3} \right]. \qquad (7)$$

The factor $f_i$ represents the weight fraction of the element under study, K or Si in our case. Equation (6) can now be rewritten in a differential form, as a function of nuclear recoil energy:

$$\rho_i^{4\pi} \left[ \frac{\text{pits}}{\text{cm}^2 \text{ keV}} \right] = \frac{10^{-4}}{2} n_i(T) \left\{ R_a(T) \left( 1 - \left( \frac{2.25}{S_n(T)} \right)^2 \right) + V_\perp t \left( 1 - \frac{2.25}{S_n(T)} \right) \right\} \qquad (8)$$

(where we have substituted the aforementioned values of $V_t$ and $V_\perp$ into Eq. (2)). $R_a$ is expressed in μm, t in hours and the stopping power in GeV cm$^2$/g . Only values of T larger than the threshold (~2.25 GeV cm$^2$/g) are considered.

It is possible to compute Eq. (8) with the mean values of $R_a(T)$ and $S_n(T)$ predicted by TRIM; however, individual recoil ions suffer longitudinal straggling varying the range of each track. To account for this, and in order to obtain the surface density as a function of the depth of the pit only, one must perform the following simulation: for each bin in T of Eq. (8), a large number of events proportional to its magnitude are distributed in $R_a$ according to their longitudinal straggling predicted by TRIM. For simplicity, the lateral straggling is neglected, taking all trajectories to be straight. The error so introduced is not large, since the lateral straggling is only 20% - 30% of $R_a$ for the recoil energies involved. Each event generates a pit, since the derivation of Eq. (6) already accounted for the necessary condition that $\theta > \theta_c$. For $t > R_a / V_t$, the depth of each pit is then randomized in the interval of possible values, $\zeta \in [0, R_a (1 - \sin \theta_c (T))]$ (for shorter times, this becomes $\zeta \in [0, t(V_t - V_\perp)]$). When renormalized to the number of events in the simulation, this procedure gives the desired surface density of etch-pits as a function of $\zeta$.

Figure (2) superimposes the results of the above calculation for heavy Dirac neutrinos, over typical densities of α-recoil and α-interaction pits in muscovite mica [13]. We use the conservative value of $t = 4$ h to ensure that the prolonged-etching factor in Eq. (8) is preserved. However, the WIMP density of the galactic halo used here ($\rho_{halo} = 0.4$ GeV/c$^2$/cm$^3$) may soon be proven to be optimistic if MACHO events are confirmed [3]. The fission-track age of the crystal is taken to be A = $10^9$ y. If Eq. (1) is correct (broken lines), the CDM signal is embedded in the α-recoil background. Adoption of Eq. (3) gives a distinct and larger signal with a sharp cut-off at $\zeta \sim 0.08$ μm. In either scenario, the contribution from K recoils dominates and Si recoils are negligible.

The large dependence of these results on the precise form of Eq. (1) suggests the importance of obtaining a more reliable expression before a choice of scanning technique is made; current experimental efforts are concentrating in the use of Atomic Force Microscopy (AFM), in an attempt to distinguish between α-recoils and WIMP-induced tracks in the $\zeta < 0.02$ μm region. However, the AFM scanning of the large areas required in an extensive search (~1 cm$^2$) is an arduous task under the present technology. Simpler, faster optical techniques such as Confocal Scanning Microscopy might be used instead in the more favorable scenarios pointed out. Previous reports [7,11] assumed that only heavy elements such as Fe and Cs, which are present in small concentrations in some micas, would produce enough radiation damage to leave etchable (and very shallow) tracks. It is our conclusion here that potassium recoils, even in the least favourable conditions of Eq. (1), are able to form such tracks, enhancing the utility of mica as a CDM detector. This has been recently ascertained experimentally [21].

To our knowledge, no exhaustive study has been made of the depth distribution of α-recoil pits in mica. Most authors simply acknowledge $\zeta <$ 0.02 µm. Huang and Walker [14] measured $\zeta$ for a small sample (26 pits) and found 0.007 µm < $\zeta$ < 0.015 µm, with < $\zeta$ > ~ 0.01 µm, while they acknowledge that the distribution could extend to larger depths. Given the observed constancy of $\rho_{rec}/\rho_{SF}$, one expects the magnitude of any extension of the distribution of α-recoil pits into the $\zeta$ > 0.02 µm region to be directly proportional to $\rho_{SF}$. However, this density of spontaneous fission tracks is simultaneously dependent on the age of the mica sample and on its $^{238}$U content. A true CDM signal would be totally independent of the second but directly proportional to the time of exposure (age) of the crystal to the CDM flux. Given a large collection of thermally stable mica samples with $\rho_{int}/\rho_{SF} > 0.15$, the observation of a similarly shaped distribution of pits in the range 0.02 µm < $\zeta$ < 0.08 µm, would not be evidence *per se* for CDM interactions. But if their overall surface density is better correlated to the sample's age than to its density of fission tracks, the "smoking gun" for the existence of CDM might be unveiled. A second, perhaps less direct way to differentiate CDM from α-recoil pits would be to use their different resistance to thermal fading (because of their dissimilar $S_t$ values).

**4. Conclusions.**

Fig. (3) presents the current exclusion limits from ultra-low background Germanium detector experiments [18] on heavy neutrinos of any coupling comprising a halo of density $\rho_{halo} = 0.4 \text{ GeV}/c^2/\text{cm}^3$. Displayed in the same figure are the improvements on these limits that would be obtained if less than 10 tracks / cm² are observed in the 0.02 µm < $\zeta$ < 0.08 µm region (Eq. (3) is assumed to hold). Two different etching times have been used. In an actual experiment, the linearity in t of Eq. (8) should be ascertained by multiple etching before long etching times can be used with confidence.

Two remarks are in order. First, the movement of the Earth through the halo where the CDM velocity distribution is isotropic, collimates the velocities in our reference frame, creating a preferred direction (i.e., a galactic "wind"). The orientation of any cleavage plane with respect to this preferred direction changes daily with the rotation of the Earth and is strongly dependent on geographical latitude [19]. For an ancient mica mineral, the daily rotation of the Earth, readjustments of the orientation of the strata, and especially the continental drift, guarantee that the effective CDM velocity distribution in the crystal has been isotropic, and hence any directional effects (namely a correlation between θ and T) can be neglected. Second, we concentrated on muscovite mica due to the incipient knowledge of its sensitivity to low-energy ion irradiations. However, SSNTD minerals with a higher content in heavy nuclei are more suitable for the present search, as a result of the coherent scattering cross section (i.e., the dependence on $N^2$ in Eq. (5)) and the larger stopping powers that could compensate for a smaller track-recording sensitivity. In particular, irradiations of Fe on Biotite (an iron-rich mica variety) and Zr on Zircon ($ZrSiO_4$), would be desirable to determine their possible utility.

While more precise measurements of the sensitivity of muscovite mica to potassium-induced tracks are needed, we conclude that the importance of this emerging field should not be underestimated. Arguments by Zeldovich and others [20] favor heavy Dirac neutrinos in the range $m_\delta \sim 1\,\text{TeV}/c^2$ as a CDM candidate. These are only marginally excluded by Ge detector experiments at a halo density that may be soon proved to be too optimistic. Mica searches could discover or rule out such a candidate if it plays any cosmologically important role.

ACKNOWLEDGMENTS

We are indebted to J. F. Ziegler at IBM-Research for making the TRIM code available to us, and to P.B. Price and D. Snowden-Ifft for making us aware of their more recent work. One of us (JIC) would like to thank D. L. Thornton and D. M. Vardiman at the Homestake Mining Company for helpful advice. This work was supported in part by the National Science Foundation under Grant No. PHYS-9007847.

REFERENCES
1. P. F. Smith and J. D. Lewin, Phys. Reports 187 (1990) 203.
2. J. R. Primack, D. Seckel and B. Sadoulet, Ann. Rev. Nucl. Part. Sci. 38 (1988) 751.
3. C. Alcock *et al*, Nature 365 (1993) 621; E. Aubourg *et al*, Nature 365 (1993) 623.
4. M. W. Goodman and E. Witten, Phys. Rev. D31(1985) 3059.
5. S. A. Durrani and R. K. Bull, Solid State Nuclear Track Detection (Pergamon Press, Oxford, 1987).

6. R. L. Fleischer, P. B. Price and R. M. Walker, Nuclear Tracks in Solids (California Press, Berkeley, 1975).
7. P.B. Price and D. Snowden-Ifft, Proc. 23rd International Cosmic Ray Conference - ICRC 23, Calgary, Canada / 19 - 30 Jul 1993; D. Snowden-Ifft, P.B. Price, L.A. Nagahara and A. Fujishima, Phys. Rev. Lett. 70 (1993) 2348 ; D. Snowden-ifft, Ph.D. thesis, University of California, Berkeley, 1992.
8. P. B. Price and M. H. Salamon, Phys. Rev. Lett. 56 (1986) 1226.
9. S. Orito *et al*, Phys. Rev. Lett. 66 (1991) 195.
10. J. Borg *et al* , Radiat. Eff. 65 (1982) 133; J. Borg, Ph.D. thesis, University of Paris, 1980.
11. R.V. Coleman, Q. Xue, Y. Gong and P.B. Price, Surf. Sci. 297 (1993) 359.
12. G. Somogyi and S. A. Szalay, Nucl. Instr. and Meth. 109 (1973) 211; R.P.Henke and E. V. Benton, Nucl. Instr. and Meth. 97 (1971) 483.
13. P. B. Price and M. H. Salamon, Nature 320 (1986) 425.
14. W. H. Huang and R. M. Walker, Science 155 (1967) 1103.
15. J. F. Ziegler, J. P. Biersack and U. Littmark, The stopping and range of ions in solids (Pergamon Press, Oxford, 1985).
16. D. P. Snowden-Ifft and P. B. Price, Phys. Lett. B288 (1992) 250.
17. S.P.Ahlen *et al*, Phys. Lett. B195 (1987) 603.
18. J. I. Collar *et al*, Nuc. Phys. B (Proc. Suppl.) 31(1993) 377; D.O.Caldwell, same proc., p. 371.
19. J. I. Collar and F. T. Avignone III, Phys. Rev. D47 (1993) 5238.
20. S. Dimopoulos *et al*, Nuc. Phys. B349 (1991) 714.
21. P.B. Price, Proc. of Inter. Conf. on Nonaccelerator Physics, Bangalore, Jan. 1994 (preprint).

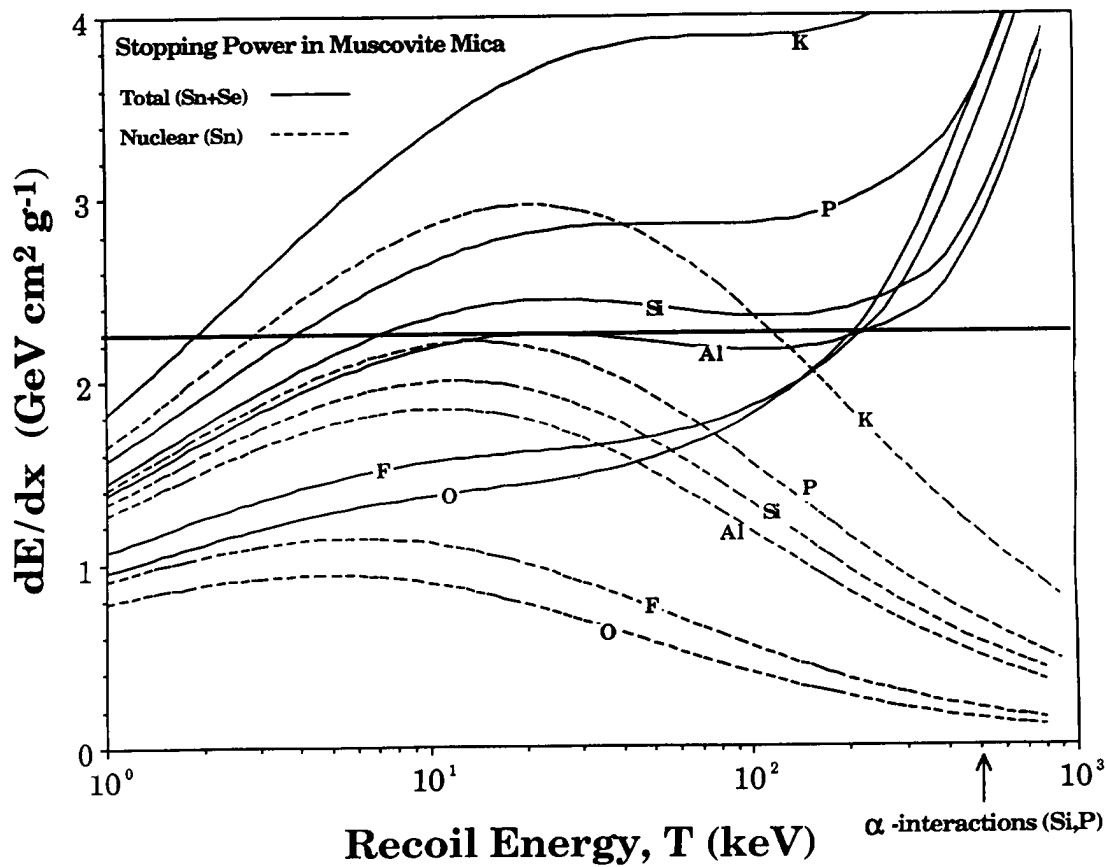

Fig. (1)
Total and nuclear stopping powers for different ions in muscovite mica, obtained from TRIM. The registration threshold for etchable tracks is indicated by the horizontal line.

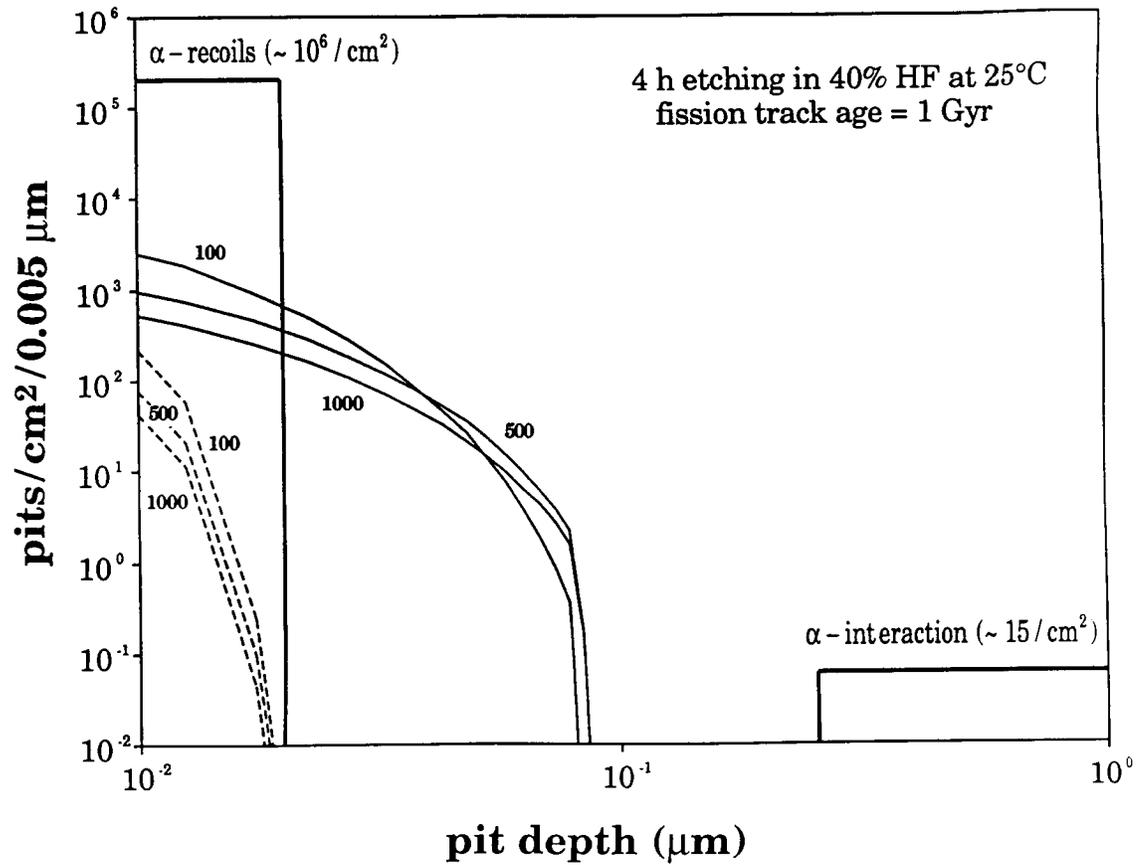

Fig. (2)
Predicted surface density of CDM induced etch-pits in muscovite mica, as a function of their depth, $\zeta$. Accepting Eq. (1), we obtain the results depicted by dashed lines. Solid lines are for the ansatz contained in Eq. (3). Lines are labeled by the mass of the heavy Dirac neutrino (taken here to be the CDM particle) in units of $GeV/c^2$. A typical surface density of other naturally occurring pits is indicated.

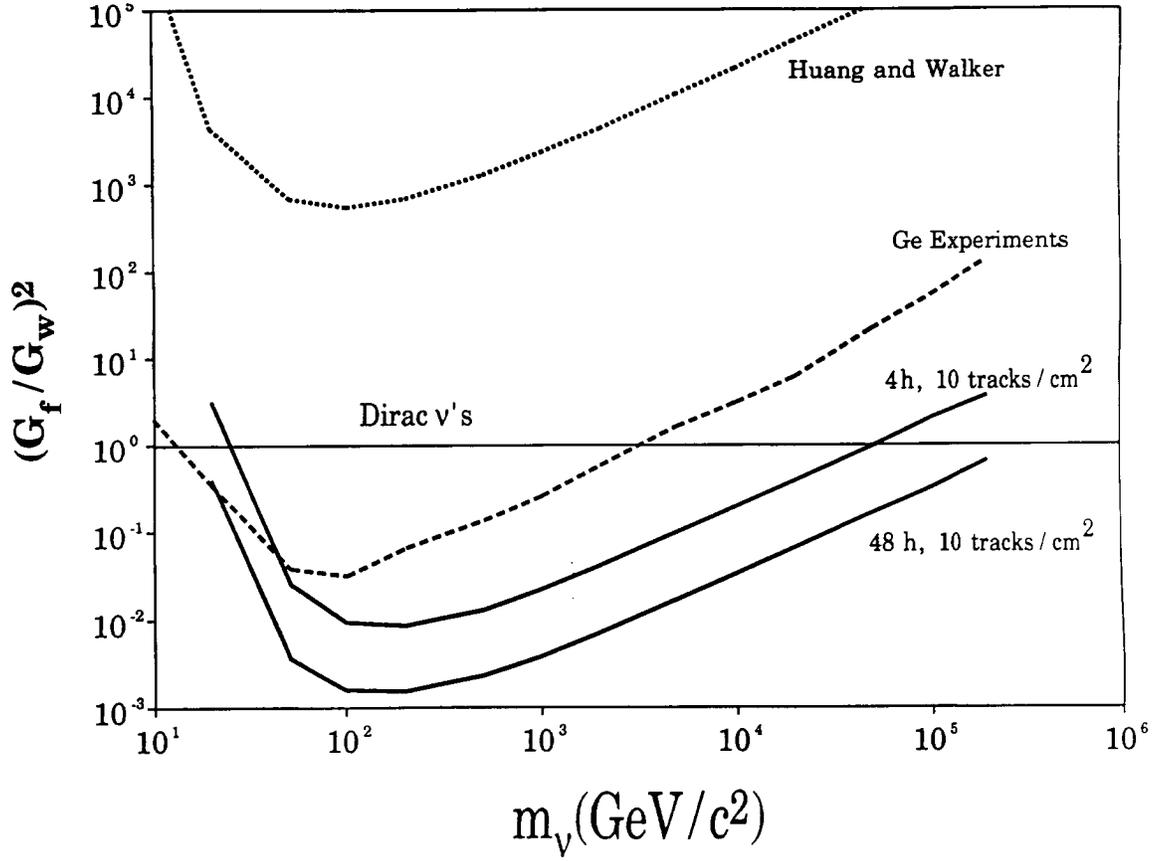

Fig. (3)
Exclusion curve from Ge experiments, and those expected if less than 10 tracks /cm² are observed in the 0.02 μm < ζ < 0.08 μm region. Heavy neutrinos with couplings above the curves are excluded at $\rho_{halo} = 0.4$ GeV/c²/cm³. By fixing the coupling constant, similar exclusion curves can be obtained in the $\rho_{halo}$ parameter (see Eq. (4)). Two different etching times are considered. The line labeled "Huang and Walker" corresponds to the modest exclusion obtained if all α-recoil pits in [14] had instead a CDM origin. A = $10^9$ y is assumed everywhere.